\documentclass[aps,prl,twocolumn,superscriptaddress,showpacs]{revtex4}
\usepackage{graphicx}

\begin{document}
\title{A variant transfer matrix method suitable for transport through multi-probe
systems}
\author{Zhenhua Qiao and Jian Wang$^*$}
\affiliation{Department of
Physics and the center of theoretical and computational physics, The
University of Hong Kong, Hong Kong, China}

\begin{abstract}
We have developed a variant transfer matrix method that is suitable
for transport through multi-probe systems. Using this method, we
have numerically studied the quantum spin-Hall effect (QSHE) on the
2D graphene with both intrinsic (${V_{so}}$) and Rashba (${V_{r}}$)
spin-orbit(SO) couplings. The integer QSHE arises in the presence of
intrinsic SO interaction and is gradually destroyed by the Rashba SO
interaction and disorder fluctuation. We have numerically determined
the phase boundaries separating integer QSHE and spin-Hall liquid.
We have found that when $V_{so} \ge 0.2t$ with $t$ is hopping
constant, the energy gap needed for the integer QSHE is the largest
satisfying $|E|<t$. For smaller $V_{so}$ the energy gap decreases
linearly. In the presence of Rashba SO interaction or disorders, the
energy gap diminishes. With Rashba SO interaction the integer QSHE
is robust at the largest energy within the energy gap while at the
smallest energy within the energy gap the integer QSHE is
insensitive to the disorders.
\end{abstract}
\pacs{
71.70.Ej,  
72.15.Rn,  
72.25.-b   
}
\maketitle

\section{Introduction}

Graphene is a 2-dimensional honeycomb lattice of single atomic
carbon layer and has a special band structures. With more and more
experimental discoveries and theoretical
predictions\cite{geim,zhang,zhang2,VPG1,peres,louie}, there is
currently a intense interest on electronic properties on the
graphene sheet. Especially the spin Hall effect(SHE) has the
potential to provide a purely electrical means to control the spins
of electron in the absence of non-ferromagnetic materials and
magnetic field\cite{sheng}. This is because the spin-orbit
interaction in the Graphene exerts a torque on the spin of electron
whose precessing leads to a spin polarized current. In a four probe
device, this spin polarized current can lead to a pure spin current
without accompanying charge current\cite{hank}. It has been proposed
by Haldane\cite{haldane} that a quantum Hall effect may exist in the
absence of magnetic field. Similarly, integer quantum spin-Hall
effect can exist on a honeycomb lattice when the intrinsic spin
orbit interaction is present\cite{sheng,kane}. In the presence of
disorder the charge conductance of mesoscopic conductors show
universal features with a universal conductance
fluctuation\cite{lee85} and the spin-Hall conductance also
fluctuates with a universal value\cite{ren} in the presence of spin
orbit interaction. The presence of disorder can also destroy the
integer quantum spin-Hall effect and quantum Hall
effect\cite{sheng1} for a Graphene system with intrinsic spin orbit
interaction\cite{sheng}. Hence it is of interest to map out the
phase diagram for the integer quantum spin-Hall effect. In this
paper, we investigate the disorder effect on the spin Hall current
for a four-probe Graphene system in the presence of intrinsic and/or
Rashba SO interactions, denoted as $V_{so}$ and $V_r$, respectively.
For such a system, the conventional transfer matrix method can not
be used. So the direct matrix inversion method must be used to
obtain the Green's function that is needed for the transport
properties.  As a result, the simulation of a multi-probe system
using the direct method is very calculational demanding.

In this paper, we developed an algorithm based on the idea of
transfer matrix that is much faster than the direct method. As an
application, we have numerically mapped out the phase diagram for a
two dimensional honeycomb lattice in the presence of the intrinsic
and/or Rashba SO interactions and disorders. When turning on the
Rashba SO interaction, we found that the energy gap needed for the
IQSHE is $|E|<t$ for $V_{so} \ge 0.2t$ and decreases linearly when
$V_{so}<0.2t$. In the presence of Rashba SO interaction, the phase
diagram $(E, V_r)$ is asymmetric about the Fermi energy. The IQSHE
is more difficult to destroy at the largest energy of the energy
gap. In the presence of disorder, the phase diagram $(E, W)$ is
again asymmetric about the Fermi energy but it is the smallest
energy of the energy gap that is robust against the disorder
fluctuation.

\section{theoretical formalism}

In the tight-binding representation, the Hamiltonian for the 2D
honeycomb lattice of the graphene structure can be written
as\cite{haldane,sheng}:
\begin{eqnarray}
H&=&-t
\sum_{<{ij}>}c^\dagger_{i}c_{j}+\frac{2i}{\sqrt{3}}V_{so}\sum_{\ll{ij}\gg}
{c^\dagger_{i}{\sigma}{\cdot}(\mathbf{d}_{kj}{\times}\mathbf{d}_{ik})c_{j}} \nonumber \\
&+&{i} V_{r}\sum_{<{ij}>}{ c^\dagger_{i}
\hat{\mathbf{e}}_{z}{\cdot}(\sigma{\times}{\mathbf{d}}_{ij})c_{j}}+\sum_{i}{\epsilon_{i}
c^\dagger_{i} c_{i}}
\end{eqnarray}
where $c^\dag_{i}$($c_{i}$) is electron creation
(annihilation) operator and $\sigma$ are Pauli matrices. The first
term is due to the nearest hopping. The second term is the intrinsic
spin-orbit interaction that involves the next nearest sites. Here $i$ and $j$
are two next nearest neighbor sites, $k$ is the common nearest
neighbor of $i$ and $j$, and ${\mathbf{d}}_{ik}$ describes a vector
pointing from $k$ to $i$. The third term is due to the Rashba spin-orbit
coupling. The last term is the on-site energy where $\epsilon_{i}$ is a
random on-site potential uniformly distributed in the interval $[-W/2,W/2]$.
In this Hamiltonian, we have set the lattice constant to be unity.

We consider a four-probe device as shown schematically in FIG.1a.
The four probes are exactly the extension from the central
scattering region, i.e., the probes are graphene ribbons. The number
of sites in the scattering region is denoted as $N=n_x{\times}n_y$,
where there are $n_x=8{\times}n+1$ sites on $n_y=4{\times}n$ chains
(FIG.1a shows the cell for $n=1$)\cite{foot2}. We apply external
bias voltages $V_i$ with $i=1,2,3,4$ at the four different probes as
$V_{i}=(v/2,0,-v/2,0)$. In the presence of Rashba SO interaction,
the spin is not a good quantum number. As a result, the spin current
is not conserved using the conventional definition. Hence we switch
off the Rashba SO interaction in the 2nd and 4th probes. Similar to
the setup of Ref.\cite{sheng} our setup can generate integer quantum
spin Hall effect. The difference between the setup of
Ref.\cite{sheng} and ours is that the lead in Ref.\cite{sheng} is a
square lattice without SO interactions while our lead is still
honeycomb lattice with SO interactions except that the Rashba SO
interaction has been switched off in lead 2 and 4. The use of the
square lattice as a lead has two consequences. It provides
additional interfacial scattering between scattering region and the
lead due to the lattice mismatch and the mismatch in SO
interactions. In addition, the dimension of the self-energy matrix
for the square lattice lead with SO interaction is much smaller. The
spin-Hall conductance $G_{sH}$ can be calculated from the
multi-probe Landauer-Buttiker formula\cite{hank,ren}:
\begin{eqnarray}
G_{sH}=(e/8{\pi})[(T_{2{\uparrow},1}-T_{2{\downarrow},1})-(T_{2{\uparrow},3}-T_{2{\downarrow},3})]
\end{eqnarray}
where the transmission coefficient is given by
$T_{2{\sigma,1}}=Tr(\Gamma_{2{\sigma}}G^{r}\Gamma_{1}G^{a})$ with
$G^{r,a}$ being the retarded and advanced Green functions of the
central disordered region which can be evaluated numerically. The
quantities $\Gamma_{i{\sigma}}$ are the linewidth functions
describing coupling of the probes and the scattering region and are
obtained by calculating self-energies $\Sigma^r$ due to the
semi-infinite leads using a transfer matrices method\cite{lopez84}.
In the following, our numerical data are mainly on a system with
$n=8$ or ${32\times}65$ sites in the system. To fix units,
throughout this paper, we define the Fermi-energy $E$, disorder
strength $W$, intrinsic spin-orbit coupling $V_{so}$ and Rashba
spin-orbit coupling $V_{r}$ in terms of the hopping energy $t$.

For the four-probe device, the conventional transfer matrix that is
suitable for two-probe devices can no longer be used. Below, we
provide a modified transfer matrix method for the four-probe device.
Note that the self-energy $\Sigma^r$ is a matrix with non-zero
elements at those positions corresponding to the interface sites
between a lead and the scattering region\cite{foot1}. Because
evaluating the Green's function $G^r$ corresponds to the inversion
of a matrix, a reasonable numbering scheme to the lattice sites can
minimize the bandwidth of the matrix and thus reduce the cost of
numerical computation. For example, to obtain the narrowest
bandwidth for our system we partition the system into layers shown
in FIG.1b so that there is no coupling between the next nearest
layers. We then label each site layer by layer from the center of
the system (see FIG.1a). As a result, the matrix $E-H-\Sigma^r$
becomes a block tri-diagonal matrix:
$$
E-H-\Sigma^r = \left(
\begin{array}{ccccccc}
A_1 & C_1 & . &  . & . & . \\
B_2 & A_2 & C_2 & . & . & . \\
. & . & . & . & .  & . \\
. & . & . & . & . & .  \\
. & .  & . & . & A_{m-1} & C_{m-1} \\
. & . & . & .  & B_m & A_m
\end{array}
\right)
$$
where $A_n$ is a $(128n-56) \times (128n-56)$ matrix, $C_n$ is a
$(128n-56) \times (128n+72)$ matrix, and $B_n$ is a $(128n-56)
\times (128n-184)$ matrix. Here $n=1$ corresponds to the innermost
layer and $n=m$ is for the outermost layer. A direct inversion of
this block tri-diagonal matrix is already faster than the other
labeling schemes. However, if we are interested in the transmission
coefficient, it is not necessary to invert the whole matrix. This is
because the self-energies of the leads are coupled only to $A_m$ of
the outermost layers, from Landauer-Buttiker's formula it is enough
to calculate the Green's function $G_{mm}^r$ which satisfys the
following equation,
$$
(E-H-\Sigma^r )
\left(
\begin{array}{c}
G^r_{1m} \\
G^r_{2m} \\
. \\
. \\
G^r_{m-1 m} \\
G^r_{m m}
\end{array}
\right)=\left(
\begin{array}{c}
0 \\
0 \\
. \\
. \\
0 \\
I_m
\end{array}
\right)$$ where $I_m$ is a unit matrix of dimension $m$. In general,
the solution $X_i$ of the following equation with block tri-diagonal
matrix can be easily obtained.

$$\left(
\begin{array}{cccccccccc}
A_1 & C_1 & . & .   & . & . \\
B_2 & A_2 & C_2 & .  & . & . \\
. & . & . & . & .   & . \\
. & . & . & .   & . & . \\
. & . & . & .  &  A_{m-1} & C_{m-1} \\
. & . & . & .   & B_m & A_m
\end{array}
\right) \left(
\begin{array}{c}
X_1 \\
X_2 \\
. \\
. \\
X_{m-1} \\
X_m
\end{array}
\right)=\left(
\begin{array}{c}
R_1 \\
R_2 \\
. \\
. \\
R_{m-1} \\
R_m
\end{array}
\right).$$

From the first row
$$
A_1X_1+C_1X_2=R_1,
$$
we have%
$$
X_1+A_1^{-1}C_1X_2=A_1^{-1}R_1.
$$

From the 2$^{nd}$ row,%
$$
B_2X_1+A_2X_2+C_2X_3=R_2,
$$
eliminating $X_1$, we have
$$
(A_2-B_2A_1^{-1}C_1)X_2+C_2X_3=R_2-B_2A_1^{-1}R_1.
$$

This equation can be written as%

$$
F_2X_2+C_2X_3=D_2,
$$
where
$$
F_2=A_2-B_2A_1^{-1}C_1,D_2=R_2-B_2A_1^{-1}R_1.
$$

From the 3$^{rd}$ row,
$$
B_3X_2+A_3X_3+C_3X_4=R_3,
$$
eliminating $X_2,$ we have
$$
F_3X_3+C_3X_4=D_3,
$$
where%
$$
F_3=A_3-B_3F_2^{-1}C_2,D_3=R_3-B_3F_2^{-1}D_2.
$$

Therefore, we have the following recursion relation,
$$
\begin{array}{ccc}
F_1= & A_1, & initial \\
F_i= & A_i-B_iF_{i-1}^{-1}C_{i-1}, & i=2,3,\cdots ,m \\
D_1= & R_1, & initial \\
D_i= & R_i-B_iF_{i-1}^{-1}D_{i-1}, & i=2,3,\cdots ,m
\end{array}
.
$$

Finally, we have

$$\left(
\begin{array}{cccccccccc}
F_1 & C_1 & . & .   & . & . \\
. & F_2 & C_2 & .   & . & . \\
. & . & . & . & . & . \\
. & . & .  & .  & . & . \\
. & . & .  & .  & F_{m-1} & C_{m-1} \\
. & . & .  & . & . & F_m
\end{array}
\right) \left(
\begin{array}{c}
X_1 \\
X_2 \\
. \\
. \\
X_{m-1} \\
X_m
\end{array}
\right) =\left(
\begin{array}{c}
D_1 \\
D_2 \\
. \\
. \\
D_{m-1} \\
D_m
\end{array}
\right) .$$

From the last row, we can solve for $X_m:$
$$
X_m=F_m^{-1}D_m.
$$
We can cancel $X_m$ in the last but one equation%
$$
X_{m-1}=F_{m-1}^{-1}(D_{m-1}-C_{m-1}X_m).
$$

In our case, $X_i=G^r_{i m}$ and $R_i=\delta_{i m} I_m$ and we are
only interest in the solution $G_{mm}$. Hence we have the solution
$$
G^r_{mm}=F_m^{-1}
$$
where
$$
\begin{array}{ccc}
F_1= & A_1,\\
F_i= & A_i-B_iF_{i-1}^{-1}C_{i-1}, & i=2,3,\cdots ,m \\
\end{array}
$$
To test the speed of this algorithm, we have calculated the spin
Hall conductance for the four-probe graphene system with different
system size labeled by $n$ on a matlab platform. The calculation is
done at a fixed energy and for 1000 random configurations. The cpu
times are listed in Table 1 where speed of direct matrix inversion
and the algorithm just described are compared. We see that the speed
up factor increases as the system size increases. For instance, for
$n=8$ which corresponds to 2080 sites (amounts to a $4016 \times
4016$ matrix) in the scattering region, a factor of 100 is gained in
speed. We note that in the presence of intrinsic SO interaction the
coupling involves next nearest neighbor interaction. This is the
major factor that slows down our algorithm. As shown in TABLE 1, for
a square lattice without intrinsic SO interaction but with Rashba SO
interaction, the speed up factor is around 200 for a $40 \times 40$
system (matrix dimension $3200$). The new algorithm is particular
useful when the large number of disorder samples and different
sample sizes are needed for the calculation of the conductance
fluctuation and its scaling with size. Finally, we wish to mention
that this algorithm also applies to multi-probe systems such as
six-probe systems.

\section{numerical results}
It has been shown that in the presence of disorder or Rashba SO
interaction the QSHE may be destroyed\cite{sheng}. As an application
of our algorithm, we study the phase phase boundary between regimes
of the integer QSHE regime and the QSH liquid in the presence of
disorder. For this purpose, we set a criteria for the QSH, i.e., if
$G_{sH}{\geq}0.999$ we say it reaches an integer quantum spin Hall
plateau (IQSH). Since the integer QSHE is due to the presence of
intrinsic SOI, we first study the phase diagram of a clean sample in
the absence of Rashba SOI, i.e., the two-component Haldane's
model\cite{haldane}. For this model, there is an energy gap within
which the IQSH effect exists. FIG.2 depicts the phase diagram in
($E$,$V_{so}$) plane with a curve separates the integer QSHE and SHE
liquid. We see that the phase diagram is symmetric about the Fermi
energy $E$ and the integer QSHE exists only for energy $E<1$ that
corresponds to the energy gap. FIG.2 shows that the energy gap
depends on the strength of intrinsic SO interaction. When $V_{so}
\ge 0.2$ the energy gap is the largest between $E=[-1,1]$ while for
$V_{so}<0.2$, the energy gap gradually diminishes to zero in a
linear fashion. Our numerical data show that for $V_{so}<0.025$ the
IQSHE disappears (see FIG.2). Between $V_{so}=[0.025,0.18]$ the
phase boundary is a linear curve. When $V_{so}>0.20$, the phase
boundary becomes a sharp vertical line.

For Haldane's model, the $\sigma_z$ is a good quantum number.
However, in the presence of Rashba SOI the spin experiences a spin
torque while traversing the system. This can destroy the IQSHE at
large enough Rashba SOI strength $V_r$. In FIG.3, we show the
spin-Hall conductance $G_{sH}$ vs Fermi energy at difference $V_r$
when $V_{so}=0.1,0.2$. In FIG.3a we see that when $V_{r}=0$, the
spin-Hall conductance is quantized between $E=-0.52$ and $+0.52$. As
$V_{r}$ increases to $0.1$, and the energy gap decreases to $-0.22$
and $0.51$. Upon further increasing $V_{r}$ to $0.2$ and $0.3$, the
gaps shrink to, respectively, $[0.06,0.50]$ and $[0.34,0.46]$. In
Ref.\cite{sheng} the IQSHE is completely destroyed when $V_r=0.3$
which is different from our result. The difference is due to the
lead used in Ref.\cite{sheng} that causes additional scattering. The
larger intrinsic SO interaction strength $V_{so}$, the more
difficult to destroy the integer QSHE as can be seen from FIG.3b.

In the presence of Rashba SO interaction the phase diagram in
$(E,V_r)$ plane at different intrinsic SO interaction strengths is
shown in FIG.4. We see that the phase diagram is asymmetric about
the Fermi energy and it is more difficult to destroy the integer
QSHE for largest positive energies within the energy gap, e.g., near
$E=0.51$ when $V_{so}=0.1$. Similar to FIG.2, we see that when
$V_{so}>0.2$ integer QSHE can exist for all energies as long as
$|E|<1$. Roughly speaking, the energy gap decreases linearly with
increasing of Rashba SOI and there is a threshold $V_r$ beyond which
the integer QSHE disappears. For instance, when $V_r>0.3$ and
$V_{so}=0.1$, the integer QSHE is destroyed.

From the above analysis, we see that $V_{so}=0.2$ is an important
point separating two different behaviors in $(E,V_{so})$ and
$(E,V_r)$ phase diagrams. Now we examine the effect of disorder on
the QSHE. FIG.5 shows the phase diagram of integer QSHE on $(E,W)$
at two typical intrinsic SO interaction strengths $V_{so}=0.1$ and
$V_{so}=0.2$. The phase diagrams are asymmetric about the Fermi
energy. Generally speaking, the larger the Rashba SO interaction
strength $V_{r}$, the smaller the energy gap needed for integer
QSHE. We already see from FIG.4 that the integer QSHE is more robust
against Rashba SO interaction strength $V_r$ at positive Fermi
energy within the energy gap. In contrast, it is small Fermi
energies within the energy gap that are stable against the disorder
fluctuation, especially for large Rashba SO interaction strength. In
addition, the phase boundary at positive Fermi energy are not very
sensitive to the variation of Rashba SO interaction strength. The
larger the intrinsic SO interaction, the larger the disorder
strength $W_c$ needed to destroy the integer QSHE. In FIG.6, we
estimate this critical disorder strength $W_c$ and plot it vs
$V_{so}$ for $E=0.01$ and $V_r=0$.

If we replace the Rashba SO interaction by the Dresselhaus SO
interaction, we have numerically confirmed that the phase diagram of
IQSHC in $(E,W)$ plane is the same if we change $E$ by $-E$.

In summary, we have developed variant transfer matrix method that is
suitable for multi-probe systems. With this algorithm, the speed
gained is of a factor 100 for a system of 2080 sites with the next
nearest SO interaction on a honeycomb lattice. For the square
lattice with Rashba SO interaction, the speed gained is around 200
for a $40 \times 40$ system. Using this algorithm, we have studied
the phase diagrams of the graphene with intrinsic and Rashba SO
interaction in the presence of disorder.

\bigskip

\section{acknowledgments}
This work was financially supported by RGC grant (HKU 7048/06P) from
the government SAR of Hong Kong and LuXin Energy Group. Computer
Center of The University of Hong Kong is gratefully acknowledged for
the High-Performance Computing facility.

\begin{figure}
\includegraphics[width=8cm,totalheight=12cm,angle=0]{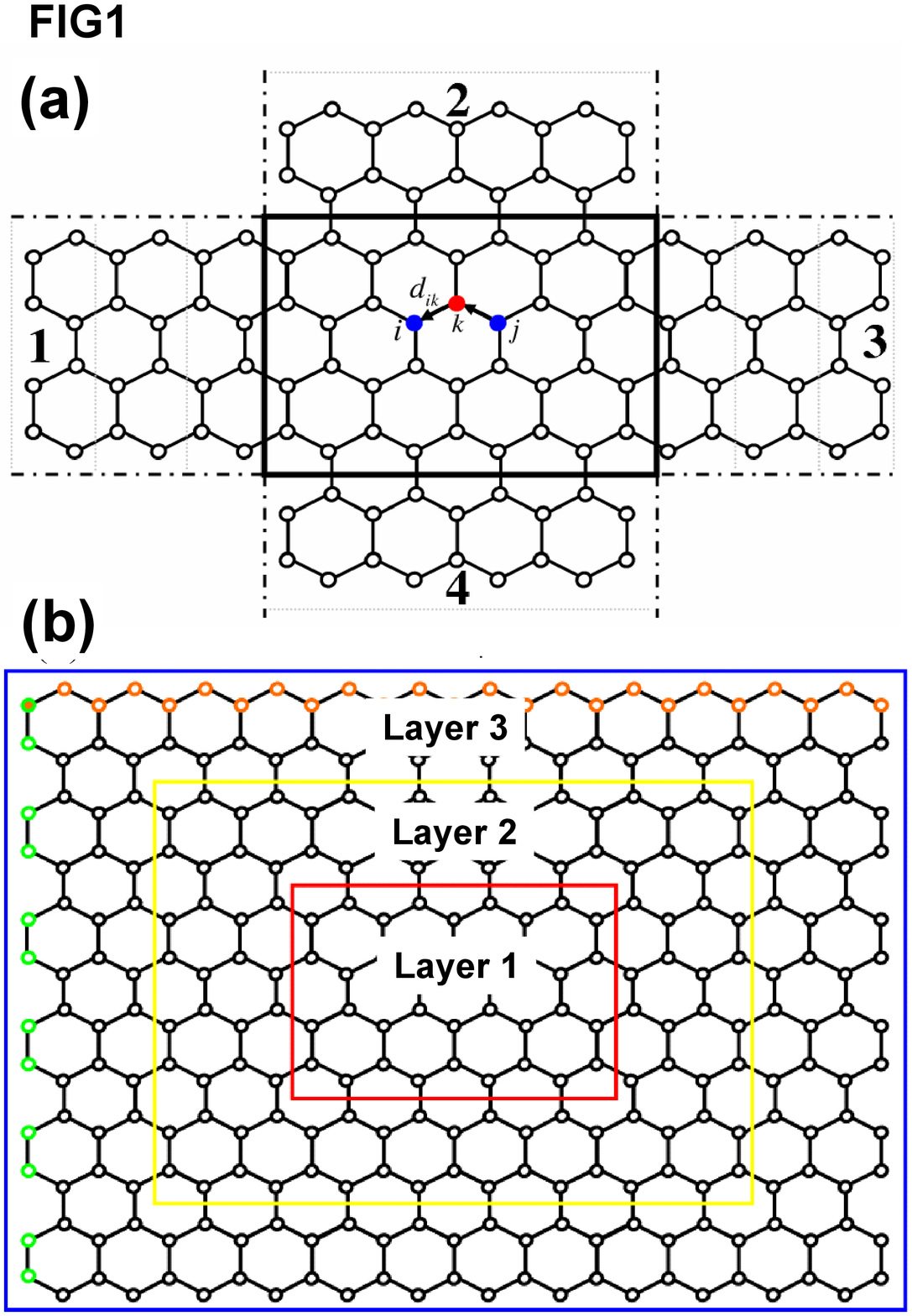}
\caption{(Color online)Schematic plot of the four terminal
mesoscopic sample where the intrinsic SO interaction exists in the
center scattering region and the leads $1$, $3$. And the Rashba SO
only exists in the center part and the leads $1$, $3$, when the
spin-Hall conductance is measured through leads $2$, $4$.}
\label{Fig.1}
\end{figure}

\begin{figure}
\includegraphics[width=6.5cm,totalheight=9cm,angle=270]{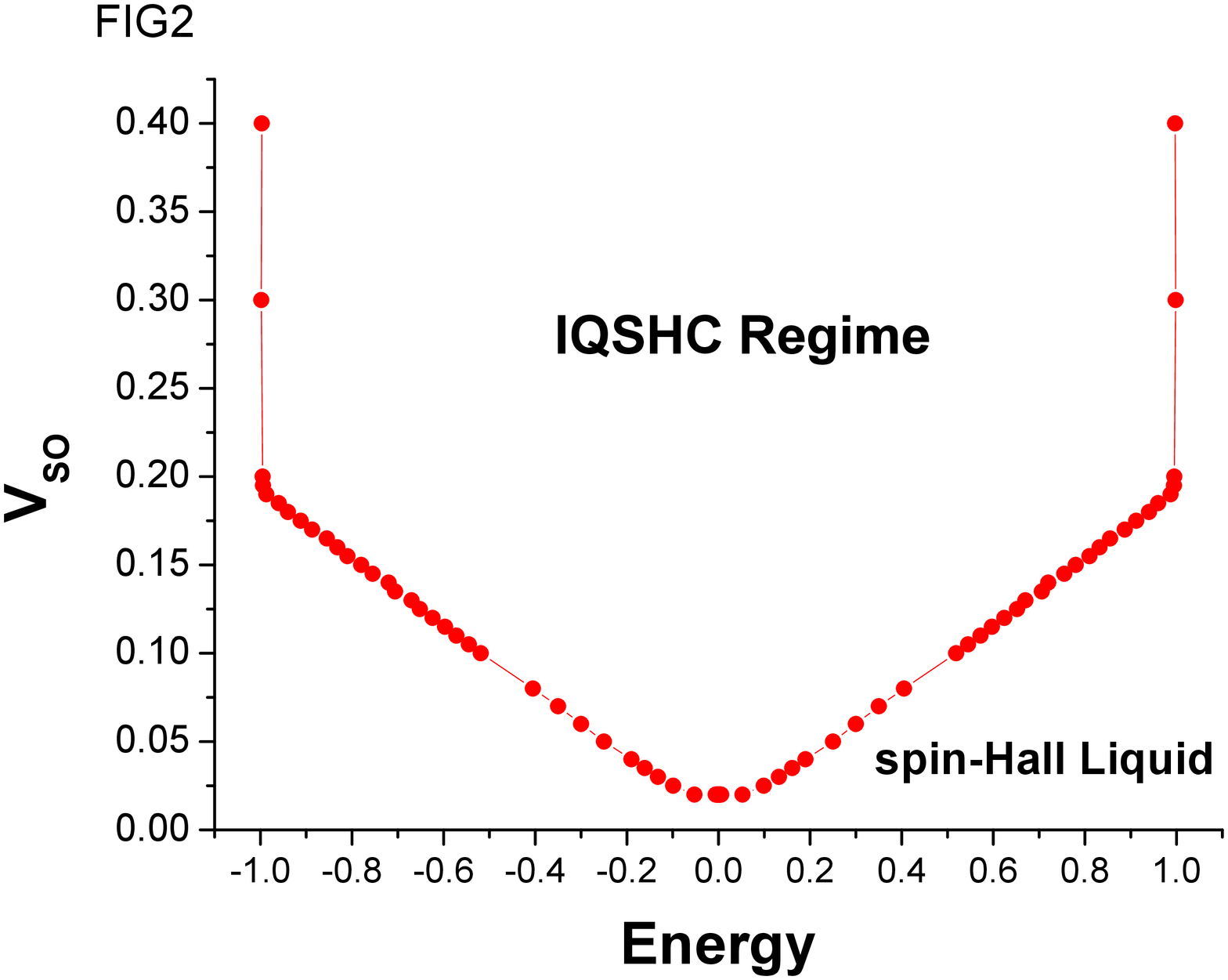}
\caption{(Color online)Phase diagram of IQSHC on ($E$, $V_{so}$)
plane for $W=0$ and $V_{r}=0$. The curve separates the IQSHC regime
and the spin-Hall liquid regime.} \label{Fig.2}
\end{figure}

\begin{figure}
\includegraphics[width=7.5cm,totalheight=9.5cm,angle=270]{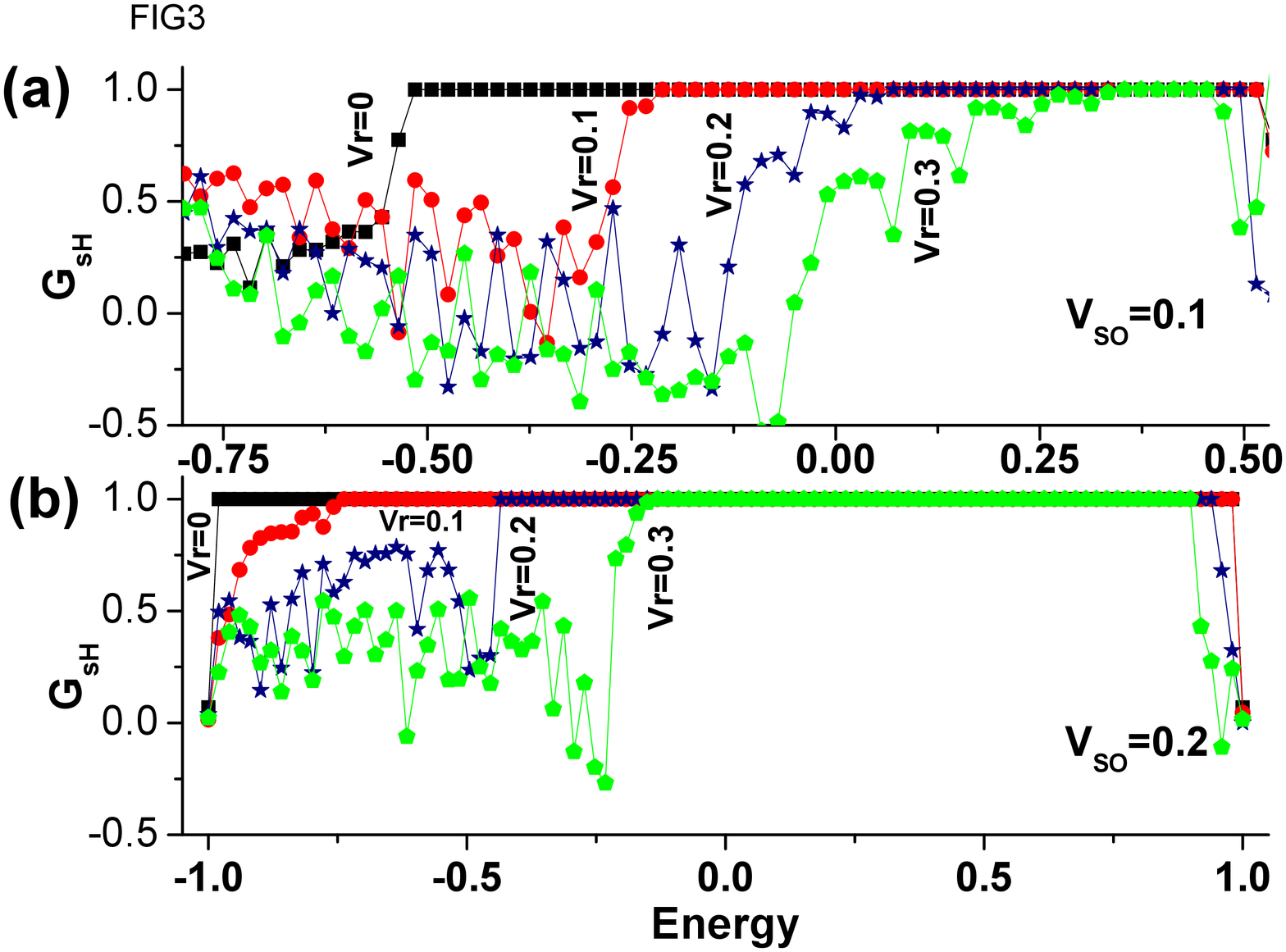}
\caption{(Color online)spin Hall Conductance versus electron Fermi
energy for $V_{r}$=0, 0.1, 0.2, 0.3 on the $N=32{\times}65$ sample.
(a) for $W=0$ and $V_{so}=0.1$; (b) for $W=0$ and $V_{so}=0.2$.}
\label{Fig.3}
\end{figure}

\begin{figure}
\includegraphics[width=5.5cm,totalheight=8cm,angle=90]{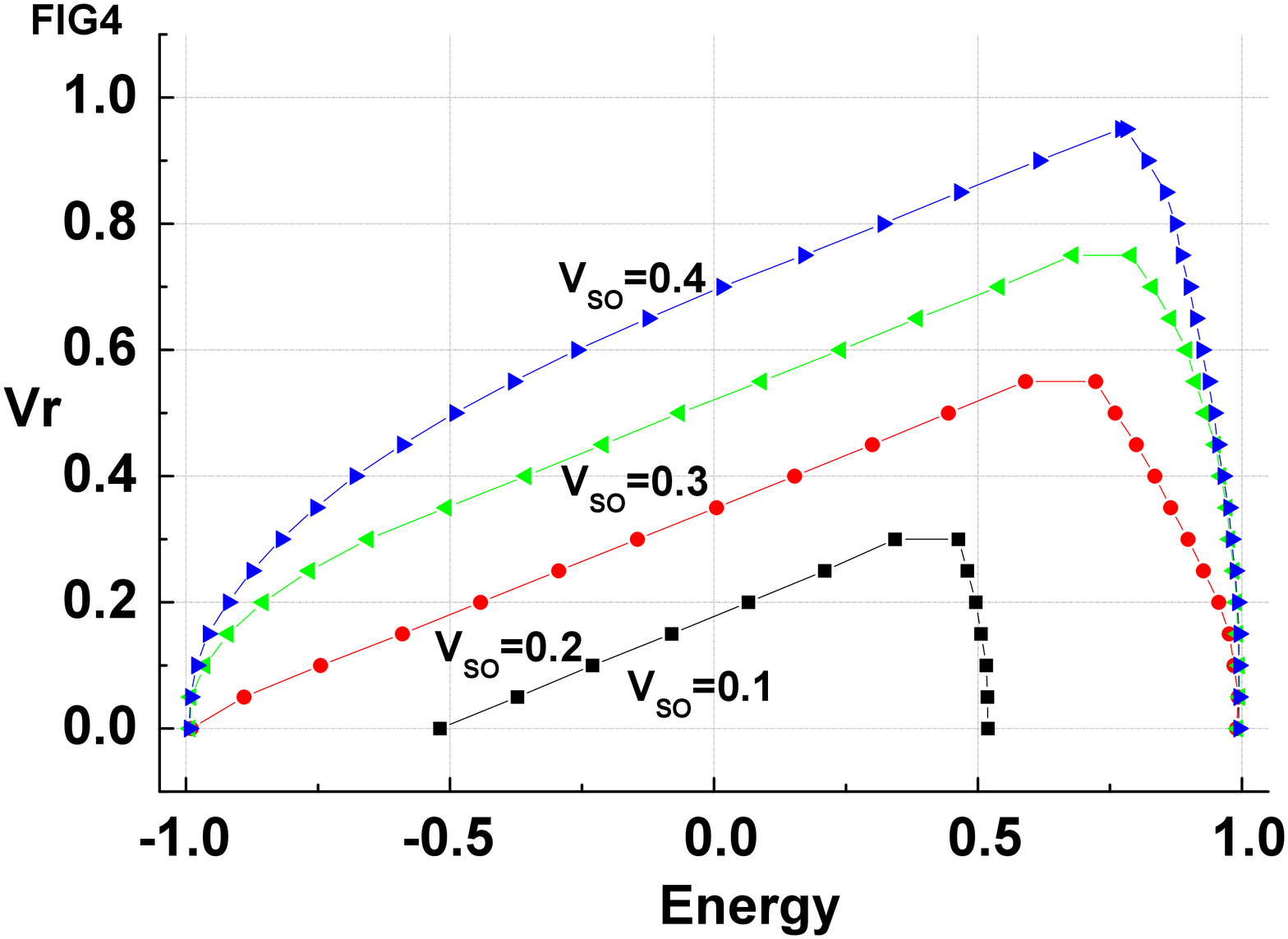}
\caption{(Color online)Phase diagram for integer quantum spin Hall
conductance on ($E$,$V_{r}$)plane. Squares, circles, left-triangles
and right-triangles are for $V_{so}$=0.1, 0.2, 0.3, 0.4,
respectively. The areas encircled by the curves and the $V_{r}$=0
line are the integer quantum spin Hall conductance regimes for
different intrinsic SOI.} \label{Fig.4}
\end{figure}

\begin{figure}
\includegraphics[width=7.0cm,totalheight=9cm,angle=270]{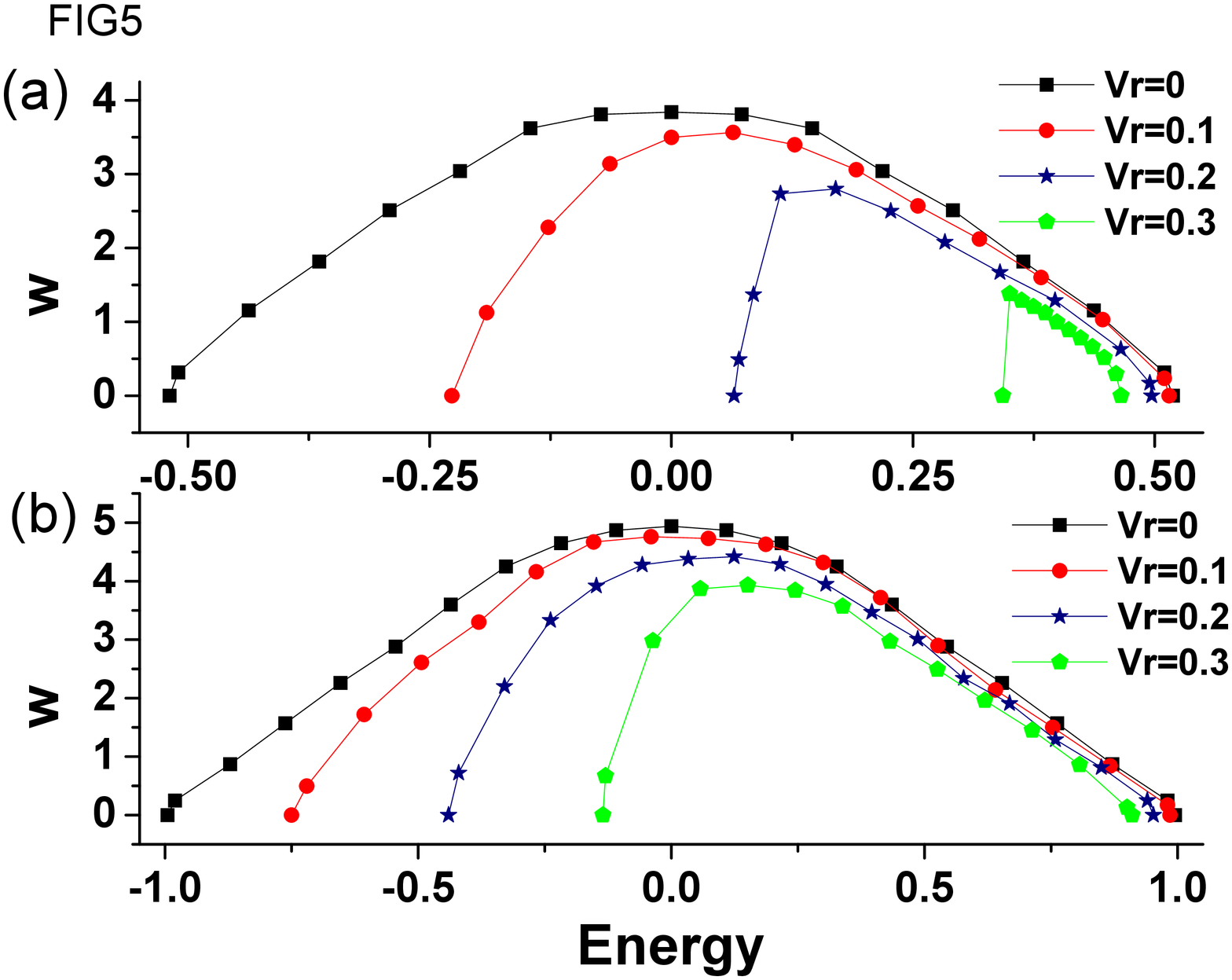}
\caption{(Color online)Phase diagram of IQSHC on ($E$, $W$) plane
for different Rashba SO coupling in the presence of (a) $V_{so}=0.1$
and (b)$V_{so}=0.2$. Squares, circles, stars and rumbus are for
$V_{r}$=0, 0.1, 0.2, 0.3. The areas encircled by the curves and the
$W$=0 line are the IQSHC regimes for different Rashba SOI.}
\label{Fig.5}
\end{figure}

\begin{figure}
\includegraphics[width=6cm,totalheight=9cm,angle=270]{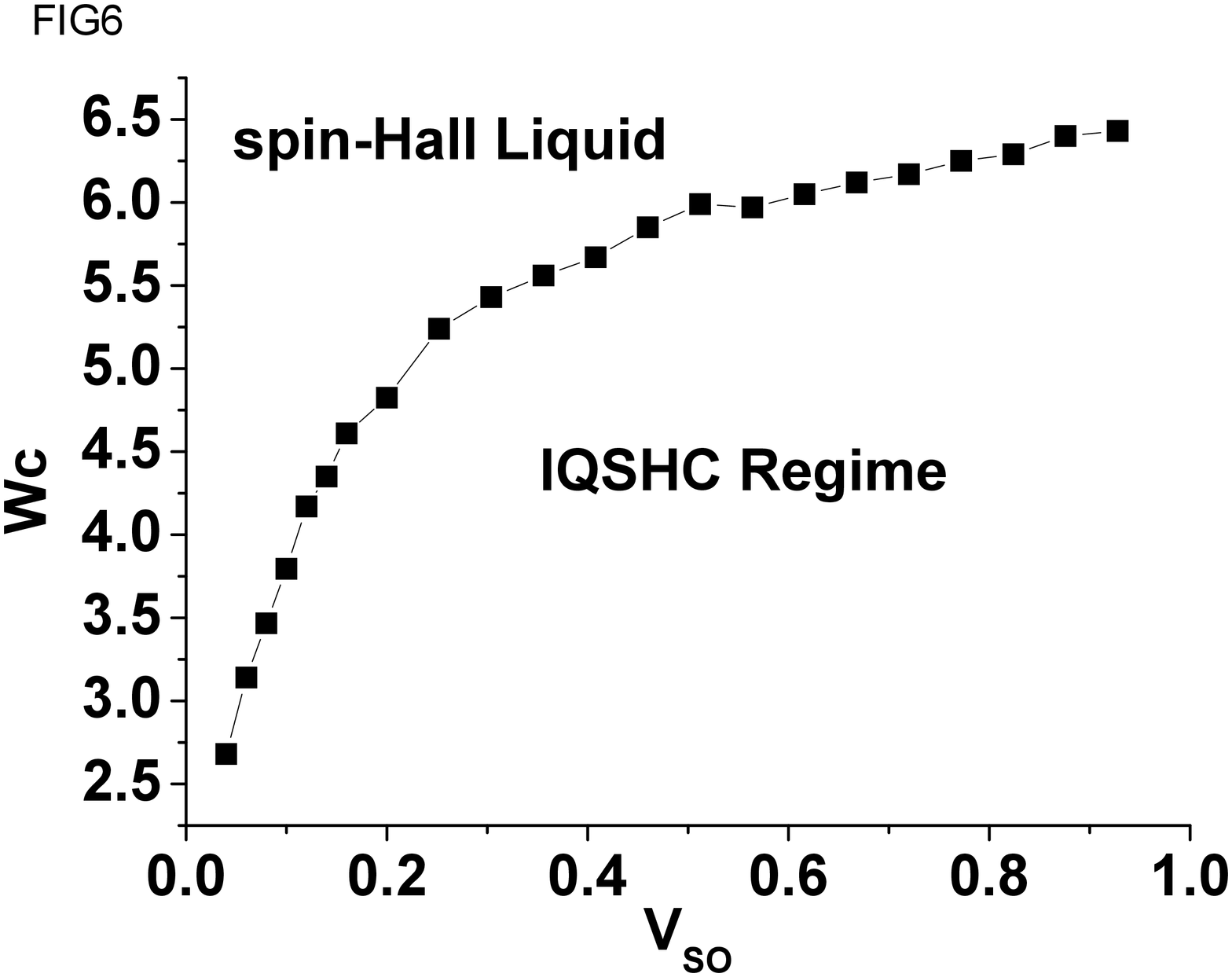}
\caption{The critical disorder strength versus intrinsic SO coupling
 $V_{so}$. The corresponding Fermi energy is $E=0.01$ and
 $V_{r}=0$. The spin Hall conductance in the regime encircled by
 the curve and the Rashba SOI axis is well quantized.}
\label{Fig.6}
\end{figure}

\begin{widetext}
\begin{table}
\includegraphics[width=12cm,totalheight=18cm,angle=270]{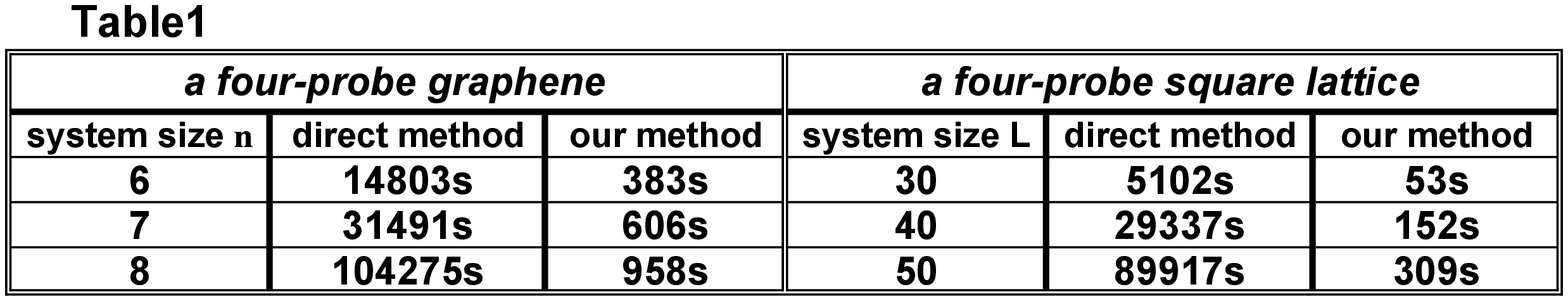}
\caption{The cpu times for different system sizes using different
methods are calculated at a fixed Fermi energy for 1000 random
configurations.} \label{TABLE.1}
\end{table}
\end{widetext}


\begin{thebibliography}{99}
\bibitem{foot2}
Here we follow the same labeling scheme as that of
Ref.\onlinecite{sheng}.
\end{thebibliography}
\end{document}